\documentclass[%
 reprint,
showpacs,preprintnumbers,
nofootinbib,
 amsmath,amssymb,
 aps,
 showkeys,
 showpacs
]{revtex4-1}
\usepackage{graphics}
\usepackage{bbold}
\usepackage{color}

\usepackage{array}
\usepackage{booktabs}
\usepackage{amsmath}
\usepackage{graphicx}
\usepackage{subfigure}

\usepackage{amsmath}
\usepackage{slashed}
\usepackage{textpos}
\usepackage{flafter}
\usepackage{float}

\def\be		{\begin{eqnarray}}
\def\en		{\end{eqnarray}}
\def\nen	{\nonumber\end{eqnarray}}
\def\no		{\nonumber}
\def\lmo		{\left|}
\def\rmo		{\right|}
\def\lt		{\left(}
\def\rt		{\right)}

\def\bfr		{\begin{flushright}}
\def\efr		{\end{flushright}}

\def\hh		{\hspace{5 mm}}

\def\jp		{{\ensuremath{J\!/\psi}}}
\def\psii		{{\ensuremath{\psi(2S)}}}

\def\LL		{\ensuremath{\Lambda\overline{\Lambda}}}
\def\Ss		{\ensuremath{\Sigma^0\overline{\Sigma}{}^0}}

\def\ov		{\ensuremath{\overline}}
\def\ee		{\ensuremath{e^+e^-}}
\def\BB		{\ensuremath{B\overline{B}}}

\def\ge	    {\ensuremath{g_E}}
\def\gm	    {\ensuremath{g_M}}

\def \br {{{\mathcal B}}}

\begin{document}

\title{A model to explain angular distributions of $\jp$ and $\psii$ decays into $\Lambda \overline \Lambda$ and  $\Sigma^0 {\overline \Sigma}{}^0$}
%
%\author{Author list}
\author{
\begin{normalsize}
\begin{center}
M.~Alekseev$^{1,2}$, A.~Amoroso$^{1,2}$, R.~Baldini~Ferroli$^{3}$, I.~Balossino$^{4,5}$, M.~Bertani$^{3}$, D.~Bettoni$^{4}$, F.~Bianchi$^{1,2}$, J.~Chai$^{2}$, G.~Cibinetto$^{4}$, F.~Cossio$^{2}$, F.~De Mori$^{1,2}$, M.~Destefanis$^{1,2}$, R.~Farinelli$^{4,6}$, L.~Fava$^{7,2}$, G.~Felici$^{3}$, I.~Garzia$^{4}$, M.~Greco$^{1,2}$, L.~Lavezzi$^{2,5}$, C.~Leng$^{2}$, M.~Maggiora$^{1,2}$, A.~Mangoni$^{8,9}$, S.~Marcello$^{1,2}$, G.~Mezzadri$^{4}$, S.~Pacetti$^{8,9}$, P.~Patteri$^{3}$, A.~Rivetti$^{1,2}$, M.~Da~Rocha~Rolo$^{1,2}$, M.~Savri\'e$^{6}$, S.~Sosio$^{1,2}$, S.~Spataro$^{1,2}$, L.~Yan$^{1,2}$
\\
\vspace{0.2cm} {\it
$^{1}$ Universit\`a di Torino, I-10125, Torino, Italy\\
$^{2}$ INFN Sezione di Torino, I-10125, Torino, Italy\\
$^{3}$ INFN Laboratori Nazionali di Frascati, I-00044, Frascati, Italy\\
$^{4}$ INFN Sezione di Ferrara, I-44122, Ferrara, Italy\\
$^{5}$ Institute of High Energy Physics, Beijing 100049, People's Republic of China\\
$^{6}$ Universit\`a di Ferrara, I-44122, Ferrara, Italy\\
$^{7}$ Universit\`a del Piemonte Orientale, I-15121, Alessandria, Italy\\
$^{8}$ Universit\`a di Perugia, I-06100, Perugia, Italy\\
$^{9}$ INFN Sezione di Perugia, I-06100, Perugia, Italy\\
}\end{center}
\vspace{0.0cm}
\end{normalsize}
}

\begin{abstract}
\rm{BESIII} data show a particular angular distribution for the decay of the $J/\psi$ and $\psi(2S)$ mesons into the hyperons $\Lambda \overline \Lambda$ and $\Sigma^0 {\overline \Sigma}{}^0$. More in details the angular distribution of the decay $\psii \to \Ss$ exhibits an opposite trend with respect to that of the other three channels: $J/\psi \to \Lambda \overline \Lambda$, $J/\psi \to \Sigma^0 {\overline \Sigma}{}^0$ and $\psi(2S) \to \Lambda \overline \Lambda$. We define a model to explain the origin of this phenomenon.
\end{abstract}
\pacs{13.20.Gd, 14.20.Jn}
\maketitle
\section{Introduction}
\begin{figure}%[H] 
\begin{center}
	\includegraphics[width=0.9\columnwidth]{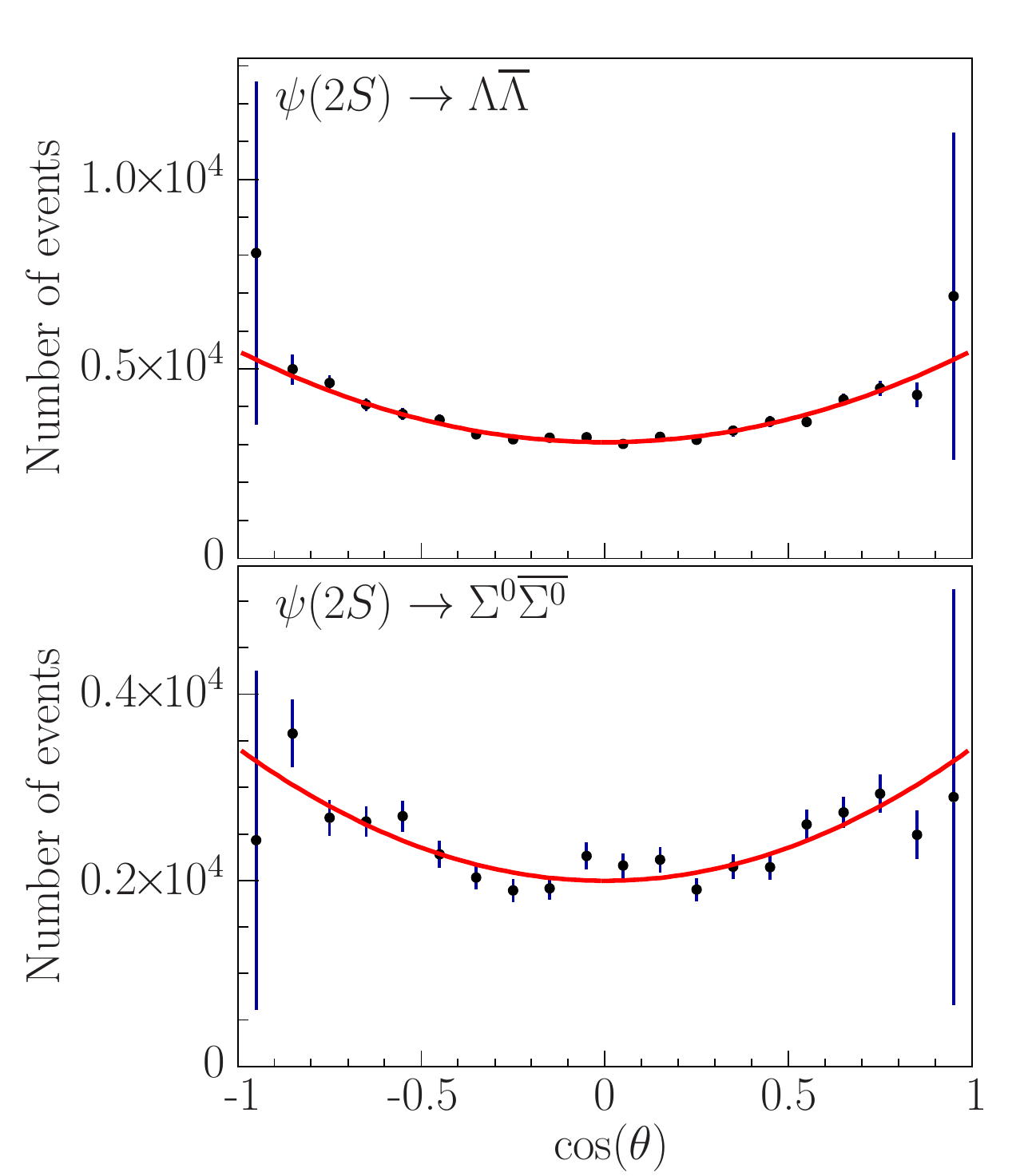}
	\caption{Angular distribution of the baryon for the $\psii$ decays into $\LL$ (upper panel) and $\Ss$ (lower panel).\label{fig.ang.distr.2s}}
\end{center}
\end{figure}

Since their discovery, charmonia, i.e., $c\ov{c}$ mesons, have been representing unique tools to deepen and expand our understanding of the strong interaction dynamics at low and medium energy ranges. Especially in case of the lightest charmonia, decay mechanisms can be studied only by means of effective models, since, due to their low-energy regime, these processes do escape the perturbative description of the quantum chromodynamics.
\\
We study the decays of the $\jp$ and $\psii$ mesons into baryon-antibaryon pairs $\BB = \LL$, $\Ss$.
The differential cross section of the process $\ee \to \psi \to \BB$ has the well known parabolic expression in $\cos\theta$~\cite{Brodsky:1981kj}
\be
\frac{dN} {d \cos \theta} \propto 1 + \alpha_B \cos^2 \theta\,,
\nen
where $\alpha_B$ is the so-called polarization parameter and $\theta$ is the baryon scattering angle, i.e., the angle between the outgoing baryon and the beam direction in the $\ee$ center of mass frame. As already pointed out in Ref.~\cite{Ablikim:2005cda}, only the decay $\jp \to \Ss$ has a negative polarization parameter $\alpha_B$. 
\begin{figure}
\begin{center}
	\includegraphics[width=0.9\columnwidth]{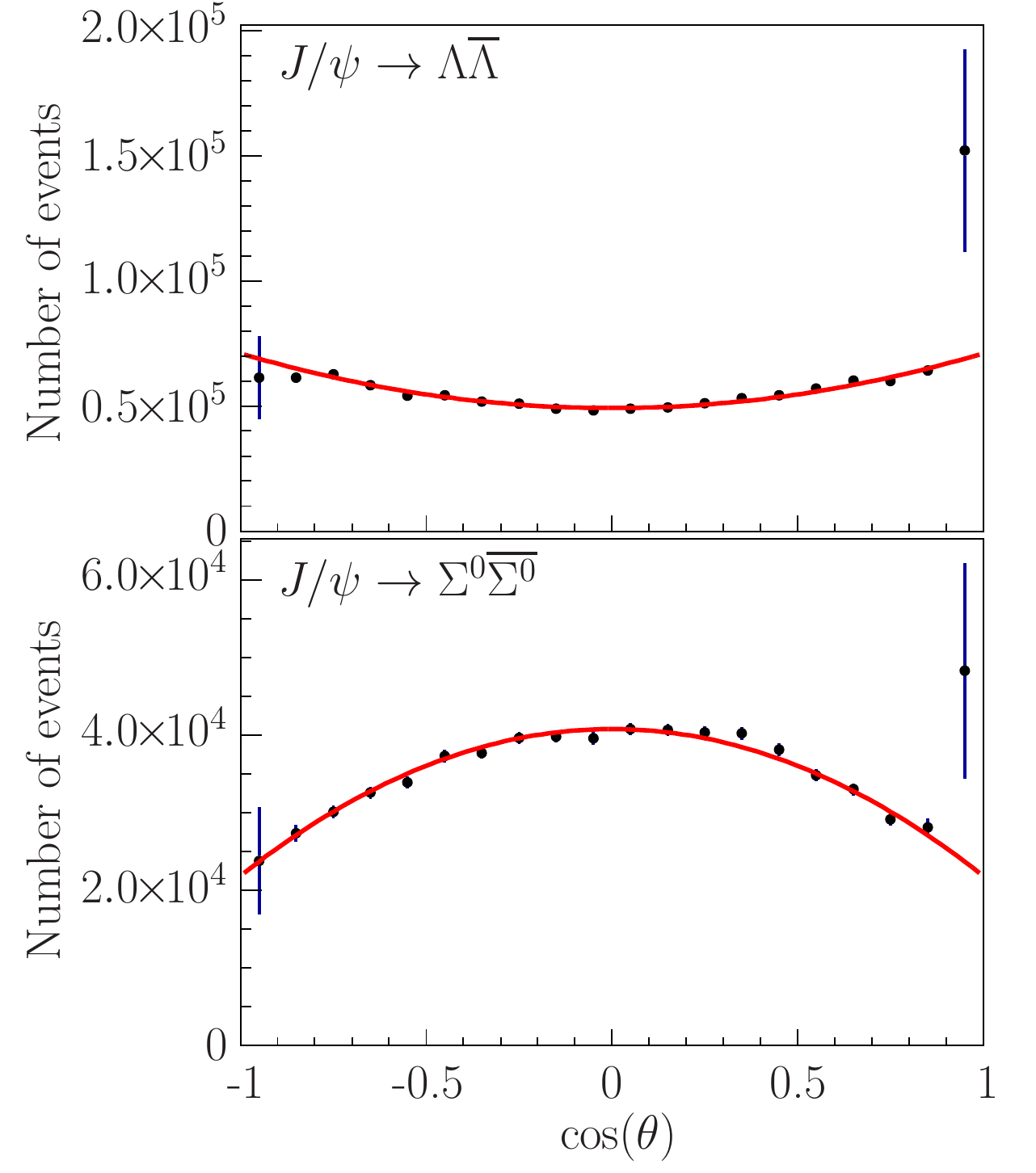}
	\caption{Angular distribution of the baryon for the $\jp$ decays into $\LL$ (upper panel) and $\Ss$ (lower panel). \label{fig.ang.distr.j}}
\end{center}
\end{figure}
Figure~\ref{fig.ang.distr.j} and~\ref{fig.ang.distr.2s} show BESIII data~\cite{brdata} on the angular distribution of the four decays: $\jp \to \LL$, $\jp \to \Ss$, and $\psii \to \LL$, $\psii \to \Ss$ respectively.
\section{Amplitudes and branching ratios}
The Feynman amplitude for the decay $\psi \to \BB$ can be written in terms of the strong magnetic and Dirac FFs as
\be
\mathcal{M}_{\psi\to\BB}=-i\epsilon_{\psi}^\mu \, 
\ov{u}(p_1)\Gamma_\mu v(p_2)
\nen
where the matrix $\Gamma_\mu$ is defined in Eq.~\eqref{eq:Gamma}, 
 $\epsilon_{\psi}^\mu$ is the polarization vector of the $\psi$ meson, and the four-momenta follow the labelling of Eq.~\eqref{eq:ee-jpsi-BB}. The branching ratio (BR) is given by the standard form for the two-body decay
\be
\mathcal{B}_{\psi\to\BB}={1 \over \Gamma_\psi}\frac{1}{8\pi}\ov{\lmo\mathcal{M}_{\psi\to\BB}\rmo^2}\frac{\lmo\vec{p}_1\rmo}{M_{\psi}^2}\,,
\nen
where $\Gamma_\psi$ is the total width of the $\psi$ meson. Using the mean value of the modulus squared of the amplitude, written in terms of the Sachs couplings,
\be
\overline{\lmo\mathcal{M}_{\psi\to\BB}\rmo^2}&\!=\!&
\frac{4}{3}M_{\psi}^2\lt
|\gm^B|^2+\frac{2M_B^2}{M_{\psi}^2}|\ge^B|^2\rt\,.
\nen
we obtain the BR
\be
\mathcal{B}_{\psi\to\BB}=\frac{M_{\psi} \beta}{12 \pi \Gamma_\psi} \lt |\gm^B|^2 + \frac{2M_B^2}{M_{\psi}^2}|\ge^B|^2 \rt .
\label{eq:BR.1}
\en
Since it does not depend on $\alpha_B$, it cannot be used to determine the polarization parameter.\\
The previous expression for the BR can be written as the sum of the moduli squared of two amplitudes 
\be
{\mathcal B}_{\psi\to\BB}=\lmo A^B_M\rmo^2+\lmo A^B_E\rmo^2\,,
\label{eq:rate.A}
\en
where, comparing with Eq.~\eqref{eq:BR.1},
\be
A^B_M=\sqrt{\frac{M_{\psi} \beta}{12 \pi \Gamma_{\psi}}}\, \gm^B
\,,\ \
A^B_E=\sqrt{\frac{M_{\psi} \beta}{6 \pi \Gamma_{\psi}}}\frac{M_B}{M_{\psi}}\, \ge^B
\,. \no
\label{eq:amp-ff}
\en
It follows that the polarization parameter of Eq.~\eqref{Eq.polpar1} can be also written as
\be
\alpha_B = {1-2|A_E^B|^2/|A_M^B|^2 \over 1+2|A_E^B|^2/|A_M^B|^2} \,.
\nen
\section{Effective model}
The SU(3) baryon octet states can be described by a matrix notation as follows~\cite{ottetto}
\be
O_B&=&\begin{pmatrix}
\Lambda/\sqrt{6}+\Sigma^0/\sqrt{2} & \Sigma^+ & p\\
\Sigma^- & \Lambda/\sqrt{6}-\Sigma^0/\sqrt{2} & n\\
\Xi^- & \Xi^0 & -2 \Lambda/\sqrt{6}\end{pmatrix} \,,
\no\\
O_{\overline{B}}&=&\begin{pmatrix}
\overline \Lambda/\sqrt{6}+\overline \Sigma^0/\sqrt{2} & \overline \Sigma^+ & \overline \Xi^+\\
\overline \Sigma^- & \overline \Lambda/\sqrt{6}- \overline \Sigma^0/\sqrt{2} & \overline \Xi^0 \\
\overline p & \overline n & -2 \overline \Lambda/\sqrt{6}\\
\end{pmatrix} \,,
\nen
where the first matrix is for baryons and the latter for antibaryons. We can consider the $J/\psi$ and $\psi(2S)$ mesons as SU(3) singlets. Following the SU(3) symmetry the level zero Lagrangian density should have the SU(3) invariant form $\mathcal L^0 \propto {\ \rm Tr}(B \overline B)$. Moreover, we consider two SU(3) breaking sources: the quark mass and the EM interaction. The first one can be parametrized by introducing the spurion matrix~\cite{Zhu:2015bha}
\be
S_m={g_m \over 3}\begin{pmatrix}
1 & 0 & 0\\
0 & 1 & 0\\
0 & 0 & -2\\
\end{pmatrix} \,,
\nen
where $g_m$ is an effective coupling constant. This matrix describes the mass breaking effect due to the mass difference between $s$ and, $u$ and $d$ quarks, where the SU(2) isospin symmetry is assumed, so that $m_u=m_d$. This SU(3) breaking is proportional to the 8$^{\rm th}$ Gell-Mann matrix $\lambda_8$. The EM breaking effect is related to the fact that the photon coupling to quarks, described by the four-current
\be
{1 \over 2} \overline q \gamma^\mu \left( \lambda_3+\lambda_8/\sqrt{3} \right) q\,,
\nen
is proportional to the electric charge. This effect can be parametrized using the following spurion matrix
\be
S_e={g_e \over 3}\begin{pmatrix}
2 & 0 & 0\\
0 & -1 & 0\\
0 & 0 & -1\\
\end{pmatrix} \,,
\nen
where $g_e$ is an EM effective coupling constant.\\
The most general SU(3) invariant effective Lagrangian density is given by~\cite{Zhu:2015bha}
\be
\mathcal L &=& g \, {\rm Tr}(OO_{\overline B}) + d \, {\rm Tr}\big(\{O, O_{\overline B} \} S_e \big)+f \, {\rm Tr}\big([OO_{\overline B} ] S_e \big) \nonumber \\
&& + d' \, {\rm Tr}\big(\{O, O_{\overline B} \} S_m \big)+f' \, {\rm Tr}\big([O, O_{\overline B} ] S_m \big) \,,
\nen
where $g$, $d$, $f$, $d'$ and $f'$ are coupling constants. 
We can extract the Lagrangians describing the $J/\psi$ and $\psi(2S)$ decays into $\Lambda \overline \Lambda$ and $\Sigma^0 \overline \Sigma^0$
\be
\label{Eq.Lagr.S.L}
{\mathcal L}_{\Ss}=(G_0+G_1) \, \Ss \,,   \
{\mathcal L}_{\Lambda \overline \Lambda} = (G_0-G_1) \, \Lambda \overline \Lambda \,,
\en
where $G_0$ and $G_1$ are combinations of coupling constants, i.e., 
\be
G_0 = g \, , \hspace{10mm} G_1 = {d \over 3} \, (2 g_m + g_e) \,.
\nen
By using the same structure of Eq.~\eqref{eq:rate.A}, the BRs 
can be expressed in terms of electric and magnetic amplitudes as
\be
\mathcal B_{\psi \to \Sigma^0 \overline \Sigma^0} &=& |A_E^\Sigma|^2 + |A_M^\Sigma|^2\,,\no\\
\mathcal B_{\psi \to \Lambda \overline \Lambda} &=& |A_E^\Lambda|^2 + |A_M^\Lambda|^2
\,.
\nen
Moreover, as obtained in Eq.~\eqref{Eq.Lagr.S.L}, such amplitudes can be further decomposed as combinations of leading, $E_0$ and $M_0$, and sub-leading terms, $E_1$ and $M_1$, with opposite relative signs, i.e.,
\be
\br_{\psi \to\Ss}&=&|E_0+E_1|^2 + |M_0+M_1|^2
\no\\
&=&|E_0|^2+|E_1|^2+2 |E_0||E_1|\cos(\rho_E) \nonumber \\
&&+ |M_0|^2+|M_1|^2+2 |M_0||M_1|\cos(\rho_M)
\no
\,,\\
\br_{\psi \to\Lambda \overline \Lambda} &=& |E_0-E_1|^2 + |M_0-M_1|^2
\no\\
&=&|E_0|^2+|E_1|^2-2 |E_0||E_1|\cos(\rho_E) \nonumber \\
&&+ |M_0|^2+|M_1|^2-2 |M_0||M_1|\cos(\rho_M) \,,
\nen
where $\rho_E$ and $\rho_M$ are the phases of the ratios $E_0/E_1$ and $M_0/M_1$.
\begin{figure}[H]
\begin{center}
\includegraphics[width=.9\columnwidth]{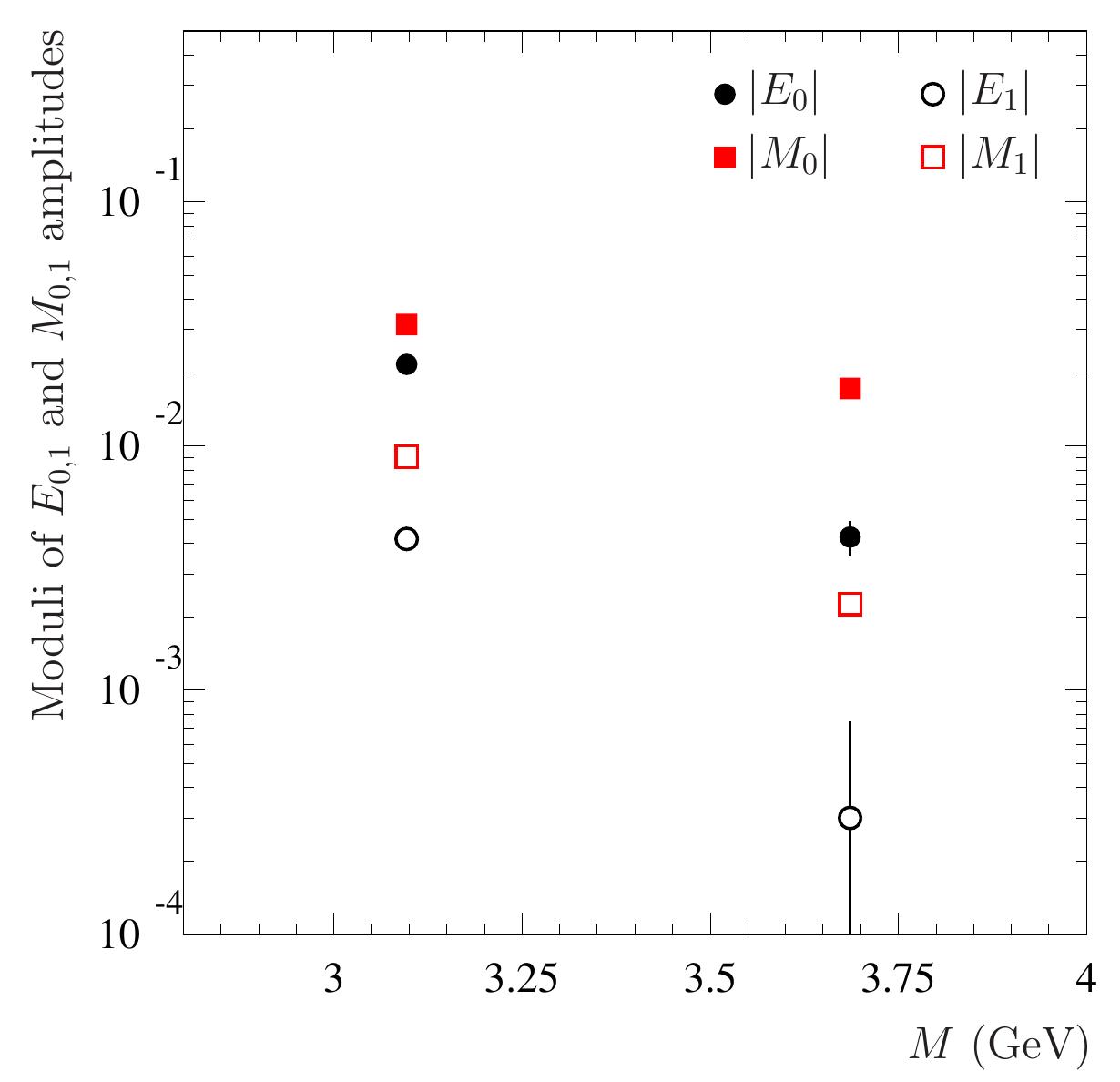}
	\caption{\label{fig:EM01} Moduli of the parameters from table~\ref{tab:results.par} as a function of the charmonium state mass $M$.}
\end{center}
\end{figure}
\begin{figure}[h]
\begin{center}
\includegraphics[width=.9\columnwidth]{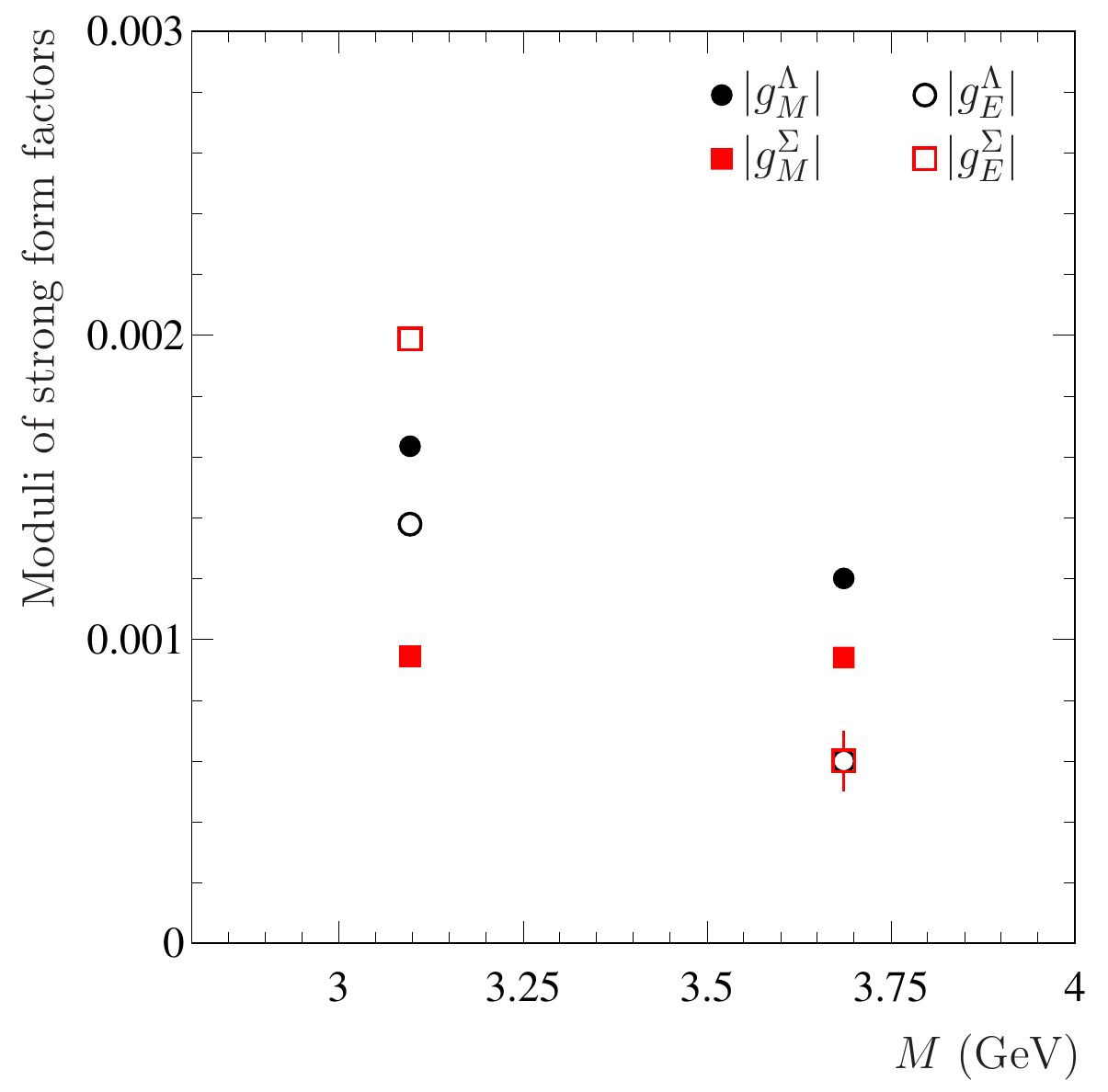}
\caption{\label{fig:gegm} Moduli of the parameters from table~\ref{tab:results.parG} as a function of the charmonium state mass $M$.}
\end{center}
\end{figure}
\begin{table}[H]
\vspace{-2mm}
\centering
\caption{Branching ratios and polarization parameters from Ref.~\cite{brdata}. In particular the value of $\alpha_B$ for the decay $\jp\to\LL$ is from Ref.~\cite{Ablikim:2018zay}.}
\label{tab:BRdata} 
\begin{tabular}{l|l|l}
\hline
Decay & BR & Pol. par. $\alpha_B$ \\
\hline
$J/\psi \to \Sigma^0 \overline \Sigma^0$ &\hfill$(11.64 \pm 0.04) \times 10^{-4}$ &\hfill$-0.449 \pm 0.020$ \\
$J/\psi \to \Lambda \overline \Lambda$ &\hfill$(19.43 \pm 0.03) \times 10^{-4}$ &\hfill$0.461 \pm 0.009$ \\
\hline
$\psi(2S) \to \Sigma^0 \overline \Sigma^0$ &\hfill$(2.44 \pm 0.03) \times 10^{-4}$ &\hfill$0.71 \pm 0.11$ \\
$\psi(2S) \to \Lambda \overline \Lambda$ &\hfill$(3.97 \pm 0.03) \times 10^{-4}$ &\hfill$0.824 \pm 0.074$ \\
\hline
\end{tabular}
\end{table}
\begin{table}[H]
\begin{center}
\caption{Moduli of the leading and sub-leading amplitudes.}
\label{tab:results.par} 
\begin{tabular}{l|l|l} 
\hline\noalign{\smallskip}
Ampl. & $J/\psi$  & $\psi(2S)$  \\
\hline
$|E_0|$ & $(2.16 \pm 0.02) \times 10^{-2}$ & $(0.42 \pm 0.07) \times 10^{-2}$ \\
$|E_1|$ & $(0.42 \pm 0.02) \times 10^{-2}$ & $(0.03 \pm 0.05) \times 10^{-2}$ \\
\hline
$|M_0|$ & $(3.15 \pm 0.02) \times 10^{-2}$ & $(1.72 \pm 0.02) \times 10^{-2}$ \\
$|M_1|$ & $(0.90 \pm 0.02) \times 10^{-2}$ & $(0.23 \pm 0.02) \times 10^{-2}$ \\
\hline
\end{tabular}
\end{center}
\end{table}
\begin{table}[H]
\begin{center}
\caption{Moduli of the strong Sachs FFs.}
\label{tab:results.parG} 
\begin{tabular}{l|l|l} 
\hline\noalign{\smallskip}
FFs & $J/\psi$  & $\psi(2S)$  \\
\hline 
$|g_E^\Sigma|$ & $(1.99 \pm 0.04) \times 10^{-3}$ &
\hfill$(0.6 \pm 0.1) \times 10^{-3}$ \\
$|g_M^\Sigma|$ &\hfill$(0.94 \pm 0.02) \times 10^{-3}$ & $(0.94 \pm 0.02) \times 10^{-3}$ \\
\hline
$|g_E^\Lambda|$ & $(1.37 \pm 0.04) \times 10^{-3}$ &\hfill$(0.6 \pm 0.1) \times 10^{-3}$ \\
$|g_M^\Lambda|$ & $(1.64 \pm 0.03) \times 10^{-3}$ &\hfill$(1.20 \pm 0.02) \times 10^{-3}$ \\
\hline
\end{tabular}
\end{center}
\end{table}
\section{Results}
\begin{figure}
\begin{flushright}
\includegraphics[width=\columnwidth]{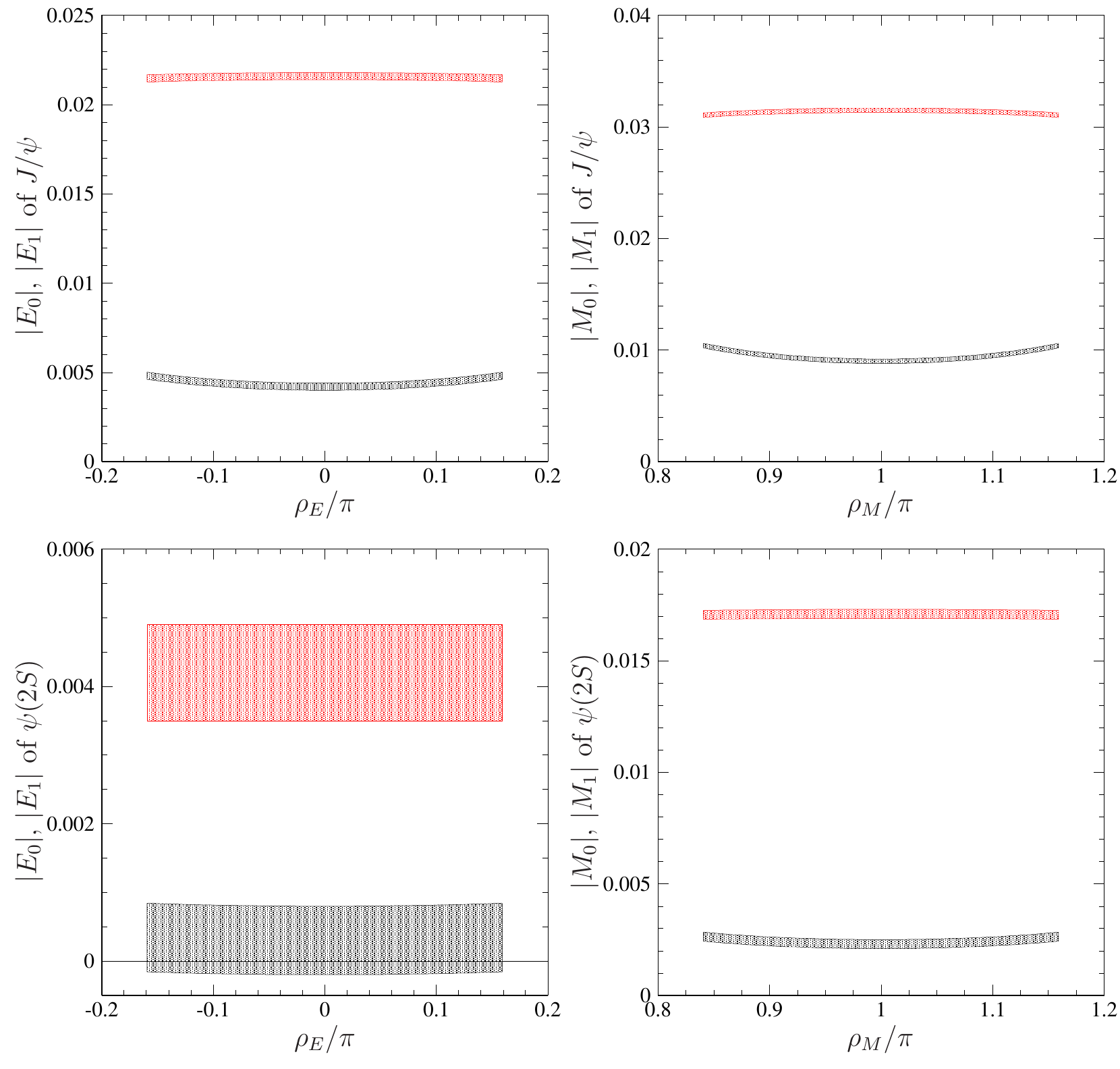}
%\put(-245,70){$R(\rho_E,\rho_M)$}
%\put(-150,10){$\rho_M$}
%\put(-170,120){$\rho_E$}
\caption{\label{fig:E0E1-M0M1-ang} Red and black bands represent moduli of leading and sub-leading amplitudes respectively. The vertical width indicates the error. Top left: moduli of amplitudes $E_0$ and $E_1$ of $J/\psi$. Top right: moduli of amplitudes $M_0$ and $M_1$ of $J/\psi$. Bottom left: moduli of amplitudes $E_0$ and $E_1$ of $\psi(2S)$. Bottom right: moduli of amplitudes $M_0$ and $M_1$ of $\psi(2S)$.}
\end{flushright}
\end{figure}

In this work we have used data from precise measurements~\cite{brdata,Ablikim:2018zay} of the BRs and polarization parameters, reported in table~\ref{tab:BRdata}, based on events collected with the BESIII detector at the BEPCII collider. 
These data are in agreement with the results of other experiments~\cite{Ablikim:2012pj,Aubert:2007uf,Pedlar:2005px,Ablikim:2006aw,Dobbs:2014ifa}. Since for each charmonium state we have six free parameters (four moduli and two relative phases) and only four constrains (two BRs and two polarization parameters), we have to fix the relative phases $\rho_E$ and $\rho_M$. 
The values $\rho_E=0$ and $\rho_M=\pi$ appear as phenomenologically favored by the data themselves. 
Indeed, (largely) different phases would give negative, and hence unphysical, values for the moduli $|E_0|$, $|E_1|$, $|M_0|$ and $|M_1|$. Moreover, as shown in Fig.~\ref{fig:E0E1-M0M1-ang}, where the four moduli for $J/\psi$ and $\psi(2S)$ are represented as functions of the phases with $\rho_E\in[-\pi/2,\pi/2]$ and $\rho_M\in[\pi/2,3\pi/2]$, the obtained results are quite stable, and the central values $\rho_E=0$, $\rho_M=\pi$ maximise the hierarchy between the moduli of leading, $E_0$ and $M_0$, and sub-leading amplitudes, $E_1$ and $M_1$.
Such values for $|E_0|$, $|E_1|$, $|M_0|$ and $|M_1|$ are reported in table~\ref{tab:results.par} and shown in Fig.~\ref{fig:EM01}. The corresponding values of $|g_{E}|$, $|g_{M}|$ are reported in table~\ref{tab:results.parG} and shown in Fig.~\ref{fig:gegm}.
The large sub-leading $\jp$ amplitudes $|E_1|$, $|M_1|$ (see table~\ref{tab:results.par} and Fig.~\ref{fig:EM01}) are responsible for the inversion of the $|\ge^B|$, $|\gm^B|$ hierarchy (see Fig.~\ref{fig:gegm} and table~\ref{tab:results.parG}).
\section{Conclusions}
Different $\Lambda$ and $\Sigma^0$ angular distributions can be explained using an effective model with the SU(3)-driven Lagrangian
\be
\mathcal{L}_{\Ss+\LL} = (G_0+G_1) \Sigma^0 \overline \Sigma^0 + (G_0-G_1) \Lambda \overline \Lambda \,.
\nen
The interplay between leading $G_0$ and sub-leading $G_1$ contributions to the decay amplitudes determines signs and values of polarization parameters $\alpha_B$.\\
In particular the different behavior of the $\jp \to \Ss$ angular distribution is due to the large values of the sub-leading amplitudes $|E_1|$ and $|M_1|$. It implies that the SU(3) mass breaking and EM effects, which are responsible for these amplitudes, play a different role in the dynamics of the $\jp$ and $\psii$ decays.\\
It is interesting to notice that angular distributions of $\Sigma^0(1385)$ and $\Sigma^\pm(1385)$, measured by BESIII~\cite{Ablikim:2016sjb,Ablikim:2016iym}, show the same $\Sigma^0$ behavior.
\\
The process $\ee\to\jp\to\Sigma^+ \overline \Sigma^-$ is currently under investigation~\cite{ref.p}, the behavior of its angular distribution would add important pieces of information to the knowledge of the \jp\ decay mechanism.
\appendix
\section{Production cross section}
\begin{figure} [H]
\begin{center}
	\includegraphics[width=.81\columnwidth]{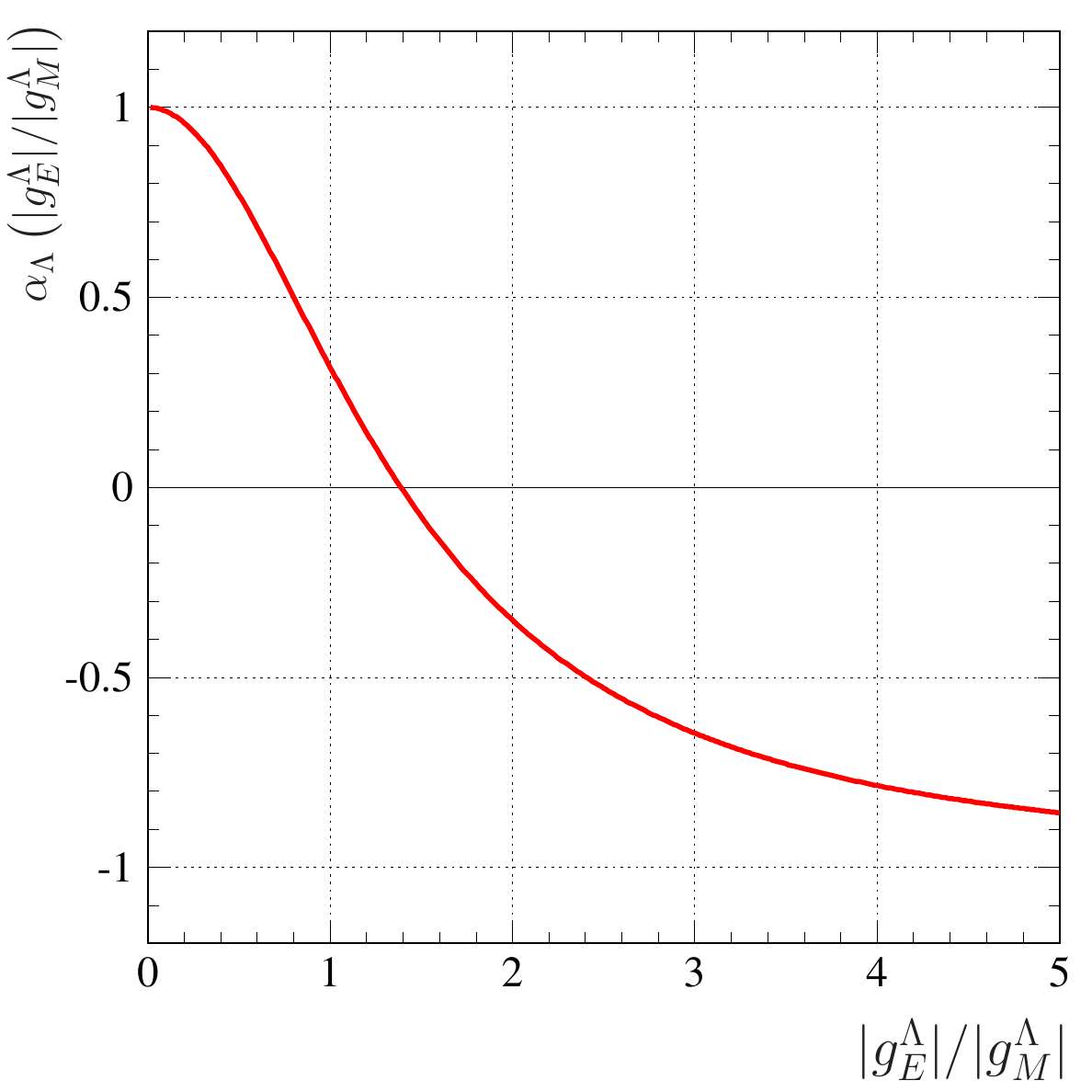}
\caption{\label{fig:pol.par.lambda.J} Polarization parameter $\alpha_\Lambda$ for the $\psi=J/\psi$ as a function of the ratio $|\ge^\Lambda|/|\gm^\Lambda|$. The masses are from Ref.~\cite{pdg}.}
\end{center}
\end{figure}

We consider the decay of a charmonium state, a $c \overline c$ vector meson $\psi$, produced via $\ee$ annihilation, into a pair baryon-antibaryon \BB, i.e., the process
\be
e^-(k_1)+ e^+(k_2)\to \psi(q)\to B(p_1)+\ov{B}(p_2) \,,
\label{eq:ee-jpsi-BB}
\en
where in parentheses are shown the 4-momenta.
\begin{figure}
	\begin{center}
		\includegraphics[width=.7\columnwidth]{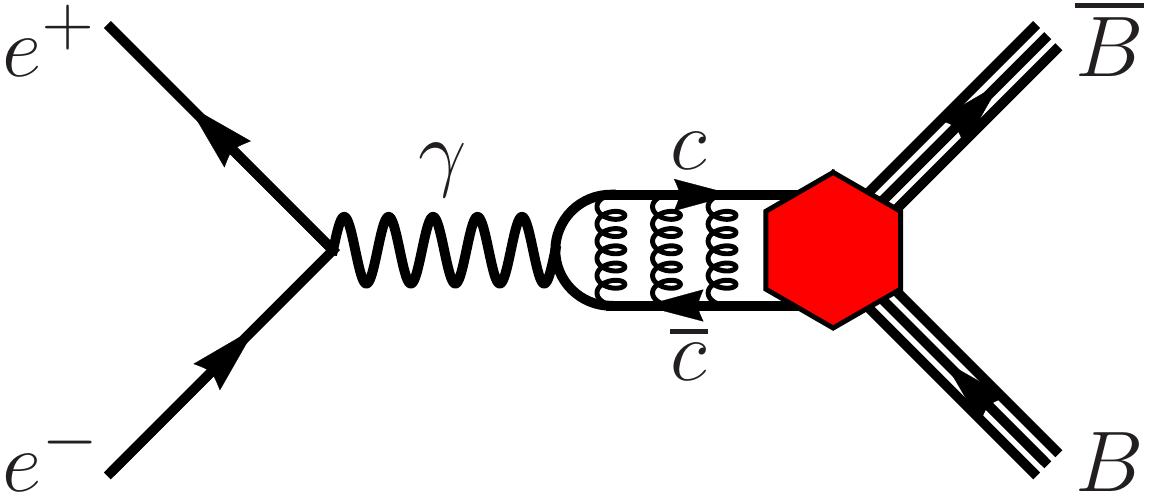}
		\caption{\label{fig:ee-jpsi-bb}Feynman diagram of the process $\ee\to\psi\to\BB$, the red hexagon represents the $\psi\BB$ coupling.}
	\end{center}
\end{figure}
The Feynman diagram is shown in Fig.~\ref{fig:ee-jpsi-bb} and the corresponding amplitude is  
\be
\mathcal{M}_{\ee\to\psi\to\BB}=-ie^2 J_B^\mu 
\ D_{\psi}\!\lt q^2\rt
\, \ov{v}(k_2)\gamma_\mu u(k_1) \,,
\nen
where $J_B^\mu=\ov{u}(p_1)\Gamma^\mu v(p_2)$ is the baryonic four-current, $D_{\psi}\lt q^2\rt$ is the $\psi$ propagator, which includes the $\gamma$-$\psi$ electromagnetic (EM) coupling, and $\ov{v}(k_2)\gamma_\mu u(k_1)$ is the leptonic four-current, the four-momenta follow the labelling of Eq.~\eqref{eq:ee-jpsi-BB}.
The $\Gamma^\mu$ matrix can be written as~\cite{dirac-pauli}
\be
\Gamma^\mu =\gamma^\mu f_{1}^B +\frac{i\sigma^{\mu\nu}q_\nu}{2M_B} \,f_2^B \,,
\label{eq:Gamma}
\en
where $M_B$ is the baryon mass and, $f_1^B$ and $f_2^B$ are constant form factors (FFs) that we call ``strong'' Dirac and Pauli couplings, they weight the vector and tensor part of the $\psi \BB$ vertex~\footnote{When the non-constant matrix $\Gamma^\mu$ is introduced to describe the EM coupling $\gamma\BB$, the tensor term contains also the anomalous magnetic moment, that, in this case where $\Gamma^\mu$ parametrises the strong vertex $\psi\BB$, has been embodied in the strong Pauli coupling.}. We can introduce the strong electric and magnetic Sachs couplings~\cite{sachs}
\be
g_E^B=f_1^B+\frac{M_{\psi}^2}{4M_B^2}f_2^B \,, \hh
g_M^B=f_1^B+f_2^B \,,
\nen
that have the same structure of the EM Sachs FFs~\cite{Claudson:1981fj}, $M_\psi$ is the mass of the charmonium state. The four quantities $f_1^B$, $f_2^B$, $\ge^B$ and $\gm^B$ are in general complex numbers. The differential cross section of the process $\ee \to \psi \to B \overline B$, in the $\ee$ center of mass frame, in terms of the two Sachs couplings, reads 
\be
\frac{d\sigma}{d\cos\theta}=
\frac{\pi\alpha^2 \beta}{2M_\psi^2} \lt \lmo\gm^B\rmo^2 + \frac{4M_B^2}{M_\psi^2}\lmo\ge^B\rmo^2 \rt \! \Big(1 + \alpha_B \cos^2 \theta \Big) \,,
\nen
where
\be
\beta=\sqrt{1 - {4 M_B^2 \over M_{\psi}^2}}
\nen
is the velocity of the out-going baryon at the $\psi$ mass, $\theta$ is the scattering angle, and the polarization parameter $\alpha_B$ is given by
\be
\label{Eq.polpar1}
\alpha_B
=\frac{M_\psi^2\lmo\gm^B\rmo^2-4M_B^2|\ge^B|^2}{M_\psi^2\lmo\gm^B\rmo^2+4M_B^2\lmo\ge^B\rmo^2} \,,
\en
$\alpha_B\in[-1,1]$. It depends only on the modulus of the ratio ${\ge^B}/{\gm^B}$, e.g., Fig.~\ref{fig:pol.par.lambda.J} shows the behavior of $\alpha_\Lambda$ in the case of $\psi=\jp$ as a function of $|g_E^{\Lambda}|/|g_M^{\Lambda}|$. The strong Sachs and, Dirac and Pauli couplings are related through $\alpha_B$. Let us consider three peculiar cases. With maximum positive polarization, $\alpha_B=1$, the strong electric Sachs coupling vanishes, i.e.,
\be
\alpha_B=1 \ \ \to \ \ g_E^B=0 \,, \ \ f_1^{B} = -\frac{M_\psi^2}{4M_B^2}f_2^{B} \,, \no \\
g_M^B=f_1^B \left( 1 -\frac{4M_B^2}{M_\psi^2} \right) = f_2^B \left( 1 -\frac{M_\psi^2}{4M_B^2} \right) \,,
\nen
the relative phase between $f_1^B$ and $f_2^B$ is $i\pi$, and the ratio of the moduli is $M_\psi^2/(4M_B^2)$.\\ 
With maximum negative polarization we have
\be
\alpha_B=-1 \ \ \to \ \ g_M^B=0 \,, \ \
f_1^{B}=-f_2^{B} \,, \no \\
g_E^B=f_1^B \left( 1 -\frac{M_\psi^2}{4M_B^2} \right) = f_2^B \left(\frac{M_\psi^2}{4M_B^2} -1 \right) \,,
\nen
in this case the strong magnetic Sachs coupling vanishes,
 the relative phase between $f_1^B$ and $f_2^B$ is $-i\pi$ and the ratio of the moduli is one.
 \\
Finally in the case with no polarization, $\alpha_B=0$, we obtain the modulus of the ratio between the Sachs couplings
\be
\alpha_B=0 \ \ \to \ \ \frac{|g_E^{B}|}{|g_M^{B}|}=\frac{M_\psi}{2M_B} \,.
\nen
\vspace{4mm}
%\clearpage
% 

\end{document}